\documentclass[10pt,letterpaper]{article}
\usepackage{opex3}
\usepackage{color}

\begin{document}

\title{High bandwidth on-chip capacitive tuning of microtoroid resonators}

\author{Christopher G. Baker$^*$, Christiaan Bekker, David L. McAuslan,  Eoin Sheridan,  and Warwick P. Bowen}

\address{Queensland Quantum Optics Laboratory, University of Queensland,
Brisbane, Queensland 4072, Australia}

\email{$^*$c.baker3@uq.edu.au} 


\begin{abstract}
We report on the design, fabrication and characterization of silica microtoroid based cavity opto-electromechanical systems (COEMS). Electrodes patterned onto the microtoroid resonators allow for rapid capacitive tuning of the optical whispering gallery mode resonances while maintaining their ultrahigh quality factor, enabling applications such as efficient radio to optical frequency conversion, optical routing and switching applications. 
\end{abstract}



\bibliographystyle{osajnl}
\bibliography{references}

\begin{thebibliography}{10}
\newcommand{\enquote}[1]{``#1''}

\bibitem{anetsberger_ultralow-dissipation_2008}
G.~Anetsberger, R.~Rivi\`{e}re, A.~Schliesser, O.~Arcizet, and T.~J.
  Kippenberg, \enquote{Ultralow-dissipation optomechanical resonators on a
  chip,} Nature Photonics \textbf{2}, 627--633 (2008).

\bibitem{rosenberg_static_2009}
J.~Rosenberg, Q.~Lin, and O.~Painter, \enquote{Static and dynamic wavelength
  routing via the gradient optical force,} Nat Photon \textbf{3}, 478--483
  (2009).

\bibitem{harris_laser_2016}
G.~I. Harris, D.~L. McAuslan, E.~Sheridan, Y.~Sachkou, C.~Baker, and W.~P.
  Bowen, \enquote{Laser cooling and control of excitations in superfluid
  helium,} Nature Physics \textbf{advance online publication} (2016).

\bibitem{verhagen_quantum-coherent_2012}
E.~Verhagen, S.~Deléglise, S.~Weis, A.~Schliesser, and T.~J. Kippenberg,
  \enquote{Quantum-coherent coupling of a mechanical oscillator to an optical
  cavity mode,} Nature \textbf{482}, 63--67 (2012).

\bibitem{gil-santos_high-frequency_2015}
E.~Gil-Santos, C.~Baker, D.~T. Nguyen, W.~Hease, C.~Gomez, A.~Lemaitre,
  S.~Ducci, G.~Leo, and I.~Favero, \enquote{High-frequency nano-optomechanical
  disk resonators in liquids,} Nature Nanotechnology \textbf{10}, 810--816
  (2015).

\bibitem{lee_cooling_2010}
K.~Lee, T.~McRae, G.~Harris, J.~Knittel, and W.~Bowen, \enquote{Cooling and
  {Control} of a {Cavity} {Optoelectromechanical} {System},} Phys. Rev. Lett.
  \textbf{104}, 123604 (2010).

\bibitem{errando2015low}
C.~Errando-Herranz, F.~Niklaus, G.~Stemme, and K.~B. Gylfason,
  \enquote{Low-power microelectromechanically tunable silicon photonic ring
  resonator add--drop filter,} Optics letters \textbf{40}, 3556--3559 (2015).

\bibitem{lee_chemically_2012}
H.~Lee, T.~Chen, J.~Li, K.~Y. Yang, S.~Jeon, O.~Painter, and K.~J. Vahala,
  \enquote{Chemically etched ultrahigh-{Q} wedge-resonator on a silicon chip,}
  Nat Photon \textbf{6}, 369--373 (2012).

\bibitem{vollmer2008whispering}
F.~Vollmer and S.~Arnold, \enquote{Whispering-gallery-mode biosensing:
  label-free detection down to single molecules,} Nature methods \textbf{5},
  591--596 (2008).

\bibitem{forstner_ultrasensitive_2014}
S.~Forstner, E.~Sheridan, J.~Knittel, C.~L. Humphreys, G.~A. Brawley,
  H.~Rubinsztein-Dunlop, and W.~P. Bowen, \enquote{Ultrasensitive
  {Optomechanical} {Magnetometry},} Adv. Mater. \textbf{26}, 6348--6353 (2014).

\bibitem{del2007optical}
P.~Del’Haye, A.~Schliesser, O.~Arcizet, T.~Wilken, R.~Holzwarth, and
  T.~Kippenberg, \enquote{Optical frequency comb generation from a monolithic
  microresonator,} Nature \textbf{450}, 1214--1217 (2007).

\bibitem{gil-santos_high-precision_2015}
E.~Gil-Santos, C.~Baker, A.~Lemaitre, C.~Gomez, S.~Ducci, G.~Leo, and
  I.~Favero, \enquote{High-precision spectral tuning of micro and nanophotonic
  cavities by resonantly enhanced photoelectrochemical etching,}
  arXiv:1511.06186 [physics]  (2015). ArXiv: 1511.06186.

\bibitem{klein_reconfigurable_2005}
E.~J. Klein, D.~H. Geuzebroek, H.~Kelderman, G.~Sengo, N.~Baker, and
  A.~Driessen, \enquote{Reconfigurable optical add-drop multiplexer using
  microring resonators,} IEEE Photonics Technology Letters \textbf{17},
  2358--2360 (2005).

\bibitem{zhang_synchronization_2012}
M.~Zhang, G.~S. Wiederhecker, S.~Manipatruni, A.~Barnard, P.~McEuen, and
  M.~Lipson, \enquote{Synchronization of {Micromechanical} {Oscillators}
  {Using} {Light},} Physical Review Letters \textbf{109}, 233906 (2012).

\bibitem{armani2003ultra}
D.~Armani, T.~Kippenberg, S.~Spillane, and K.~Vahala, \enquote{Ultra-high-q
  toroid microcavity on a chip,} Nature \textbf{421}, 925--928 (2003).

\bibitem{armani2004electrical}
D.~Armani, B.~Min, A.~Martin, and K.~J. Vahala, \enquote{Electrical
  thermo-optic tuning of ultrahigh-q microtoroid resonators,} Applied physics
  letters \textbf{85}, 5439--5441 (2004).

\bibitem{heylman_photothermal_2013}
K.~D. Heylman and R.~H. Goldsmith, \enquote{Photothermal mapping and free-space
  laser tuning of toroidal optical microcavities,} Applied Physics Letters
  \textbf{103}, 211116 (2013).

\bibitem{bogaerts_silicon_2012}
W.~Bogaerts, P.~De~Heyn, T.~Van~Vaerenbergh, K.~De~Vos, S.~Kumar~Selvaraja,
  T.~Claes, P.~Dumon, P.~Bienstman, D.~Van~Thourhout, and R.~Baets,
  \enquote{Silicon microring resonators,} Laser \& Photonics Reviews
  \textbf{6}, 47--73 (2012).

\bibitem{pollinger2009ultrahigh}
M.~P{\"o}llinger, D.~O’Shea, F.~Warken, and A.~Rauschenbeutel,
  \enquote{Ultrahigh-q tunable whispering-gallery-mode microresonator,}
  Physical review letters \textbf{103}, 053901 (2009).

\bibitem{sumetsky2010super}
M.~Sumetsky, Y.~Dulashko, and R.~Windeler, \enquote{Super free spectral range
  tunable optical microbubble resonator,} Optics letters \textbf{35},
  1866--1868 (2010).

\bibitem{yang2003fiber}
L.~Yang, D.~Armani, and K.~Vahala, \enquote{Fiber-coupled erbium microlasers on
  a chip,} Applied physics letters \textbf{83}, 825--826 (2003).

\bibitem{kippenberg2004ultralow}
T.~Kippenberg, S.~Spillane, D.~Armani, and K.~Vahala,
  \enquote{Ultralow-threshold microcavity raman laser on a microelectronic
  chip,} Optics letters \textbf{29}, 1224--1226 (2004).

\bibitem{jung_electrical_2014}
H.~Jung, K.~Y. Fong, C.~Xiong, and H.~X. Tang, \enquote{Electrical tuning and
  switching of an optical frequency comb generated in aluminum nitride
  microring resonators,} Optics Letters \textbf{39}, 84 (2014).

\bibitem{Miller:15}
S.~A. Miller, Y.~Okawachi, S.~Ramelow, K.~Luke, A.~Dutt, A.~Farsi, A.~L. Gaeta,
  and M.~Lipson, \enquote{Tunable frequency combs based on dual microring
  resonators,} Opt. Express \textbf{23}, 21527--21540 (2015).

\bibitem{baker_photoelastic_2014}
C.~Baker, W.~Hease, D.-T. Nguyen, A.~Andronico, S.~Ducci, G.~Leo, and
  I.~Favero, \enquote{Photoelastic coupling in gallium arsenide optomechanical
  disk resonators,} Opt. Express \textbf{22}, 14072--14086 (2014).

\bibitem{bowen2015quantum}
W.~P. Bowen and G.~J. Milburn, \emph{Quantum optomechanics} (CRC Press, 2015).

\bibitem{pitanti_strong_2015}
A.~Pitanti, J.~M. Fink, A.~H. Safavi-Naeini, J.~T. Hill, C.~U. Lei,
  A.~Tredicucci, and O.~Painter, \enquote{Strong opto-electro-mechanical
  coupling in a silicon photonic crystal cavity,} Opt. Express \textbf{23},
  3196--3208 (2015).

\bibitem{abdulla_tuning_2011}
S.~Abdulla, L.~Kauppinen, M.~Dijkstra, M.~de~Boer, E.~Berenschot, H.~Jansen,
  R.~de~Ridder, and G.~Krijnen, \enquote{Tuning a racetrack ring resonator by
  an integrated dielectric {MEMS} cantilever,} Optics Express \textbf{19},
  15864 (2011).

\bibitem{winger_chip-scale_2011}
M.~Winger, T.~D. Blasius, T.~P. Mayer~Alegre, A.~H. Safavi-Naeini, S.~Meenehan,
  J.~Cohen, S.~Stobbe, and O.~Painter, \enquote{A chip-scale integrated
  cavity-electro-optomechanics platform,} Opt. Express \textbf{19},
  24905--24921 (2011).

\bibitem{bagci_optical_2014}
T.~Bagci, A.~Simonsen, S.~Schmid, L.~G. Villanueva, E.~Zeuthen, J.~Appel, J.~M.
  Taylor, A.~Sørensen, K.~Usami, A.~Schliesser, and E.~S. Polzik,
  \enquote{Optical detection of radio waves through a nanomechanical
  transducer,} Nature \textbf{507}, 81--85 (2014).

\bibitem{nguyen2013ultrahigh}
D.~T. Nguyen, C.~Baker, W.~Hease, S.~Sejil, P.~Senellart, A.~Lemaitre,
  S.~Ducci, G.~Leo, and I.~Favero, \enquote{Ultrahigh q-frequency product for
  optomechanical disk resonators with a mechanical shield,} Applied Physics
  Letters \textbf{103}, 241112 (2013).

\bibitem{jiang_high-q_2009}
X.~Jiang, Q.~Lin, J.~Rosenberg, K.~Vahala, and O.~Painter, \enquote{High-{Q}
  double-disk microcavities for cavity optomechanics,} Optics Express
  \textbf{17}, 20911--20919 (2009).

\bibitem{andrews_bidirectional_2014}
R.~W. Andrews, R.~W. Peterson, T.~P. Purdy, K.~Cicak, R.~W. Simmonds, C.~A.
  Regal, and K.~W. Lehnert, \enquote{Bidirectional and efficient conversion
  between microwave and optical light,} Nature Physics \textbf{10}, 321--326
  (2014).

\bibitem{slade_electrical_2002}
P.~Slade and E.~Taylor, \enquote{Electrical breakdown in atmospheric air
  between closely spaced (0.2 μm-40 μm) electrical contacts,} IEEE
  Transactions on Components and Packaging Technologies \textbf{25}, 390--396
  (2002).

\end{thebibliography}

\section{Introduction}

High optical quality factor (Q) cavities, and in particular whispering gallery mode (WGM) resonators are broadly used optical components. Due to the combination of high Q and low mode volume, they are a platform of choice  for a wide range of applications including optomechanics \cite{anetsberger_ultralow-dissipation_2008, rosenberg_static_2009,harris_laser_2016,verhagen_quantum-coherent_2012,gil-santos_high-frequency_2015},  optoelectromechanics \cite{lee_cooling_2010}, optical add-drop filters \cite{errando2015low}, on-chip lasers \cite{lee_chemically_2012}, bio- and magnetic sensing \cite{vollmer2008whispering, forstner_ultrasensitive_2014} and frequency comb generation \cite{del2007optical}. The ability to spectrally tune the WGM resonances of these resonators, preferably in a reversible fashion, is of great interest for many of these applications. This tunability can for instance be used to overcome microfabrication induced size variability \cite{gil-santos_high-precision_2015} or temperature associated frequency drifts, lessening the need for bulky and expensive tunable laser sources. It also permits optical switching, reconfigurable optical routing \cite{klein_reconfigurable_2005} and is required for any kind of on-chip network of interacting high Q resonators \cite{zhang_synchronization_2012}.

Silica is a particularly advantageous material for on- and off-chip photonics thanks to its extremely low optical absorption, low susceptibility to two-photon absorption and associated free carrier absorption, and the absence of lossy surface states requiring passivation. In particular silica microtoroids \cite{armani2003ultra} and silica wedge resonators \cite{lee_chemically_2012} set the benchmark for the highest optical Q resonators achievable on a chip, with Q values reaching almost 1 billion \cite{lee_chemically_2012}.
However, silica resonators are difficult to tune due to the absence of free carrier injection schemes and vanishing linear electro-optic coefficient. Two main approaches have been employed thus far: heat-based and strain-based methods.
Heat-based tuning \cite{armani2004electrical, heylman_photothermal_2013} relies on changing the size and refractive index of the material with temperature, and, while allowing for large frequency shifts, suffers from several drawbacks. First, owing to silica's low thermo-optic and thermal expansion coefficients, high temperatures are required for large wavelength shifts, requiring typically tens of milliwatts of power expenditure \cite{bogaerts_silicon_2012} and precluding cryogenic or biological applications. Furthermore this tuning method is inherently slow, with modulation rates constrained to the kHz range for microtoroids \cite{armani2004electrical, heylman_photothermal_2013}.  
Strain coupling, whereby the resonator is subjected to stress and deformed, thus changing its effective size and optical resonance frequencies has been demonstrated on silica bottle resonators \cite{pollinger2009ultrahigh, sumetsky2010super} but not for integrated on-chip resonators.

Here we propose and experimentally demonstrate an efficient strain coupling approach  to tune the high Q resonances of on-chip WGM resonators, based upon capacitive actuation. Gold electrodes are patterned upon the top surface of the device to be tuned. When a bias voltage is applied, the attractive capacitive force between the electrodes strains the device and shifts the position of its optical resonances. Our approach presents several advantages.  First it allows for very fast electrical tuning of the resonances reaching up to the tens of MHz range, several orders of magnitude faster than previously demonstrated schemes, while maintaining the ultra-high Q nature of the resonances. Moreover,  capacitive tuning requires minimal  power expenditure as there is no current flow between the electrodes in the steady state. This enables dense arrays of tuneable microresonators to be combined on a single chip, with low power consumption compatible with packaged devices and without the risk of thermal crosstalk between neighbouring devices. Finally, we present numerical simulations of a new design based upon interdigitated electrodes which significantly improves upon the performance of the current devices. Our approach could have applications for silica microtoroid and microdisk based erbium lasers \cite{yang2003fiber}, Raman and Brillouin lasers \cite{kippenberg2004ultralow,lee_chemically_2012} as well as tuneable and switchable frequency combs \cite{jung_electrical_2014, Miller:15}.

\section{Device design and fabrication}
\subsection{Design}
\begin{figure}
\centering
\includegraphics[width=\textwidth]{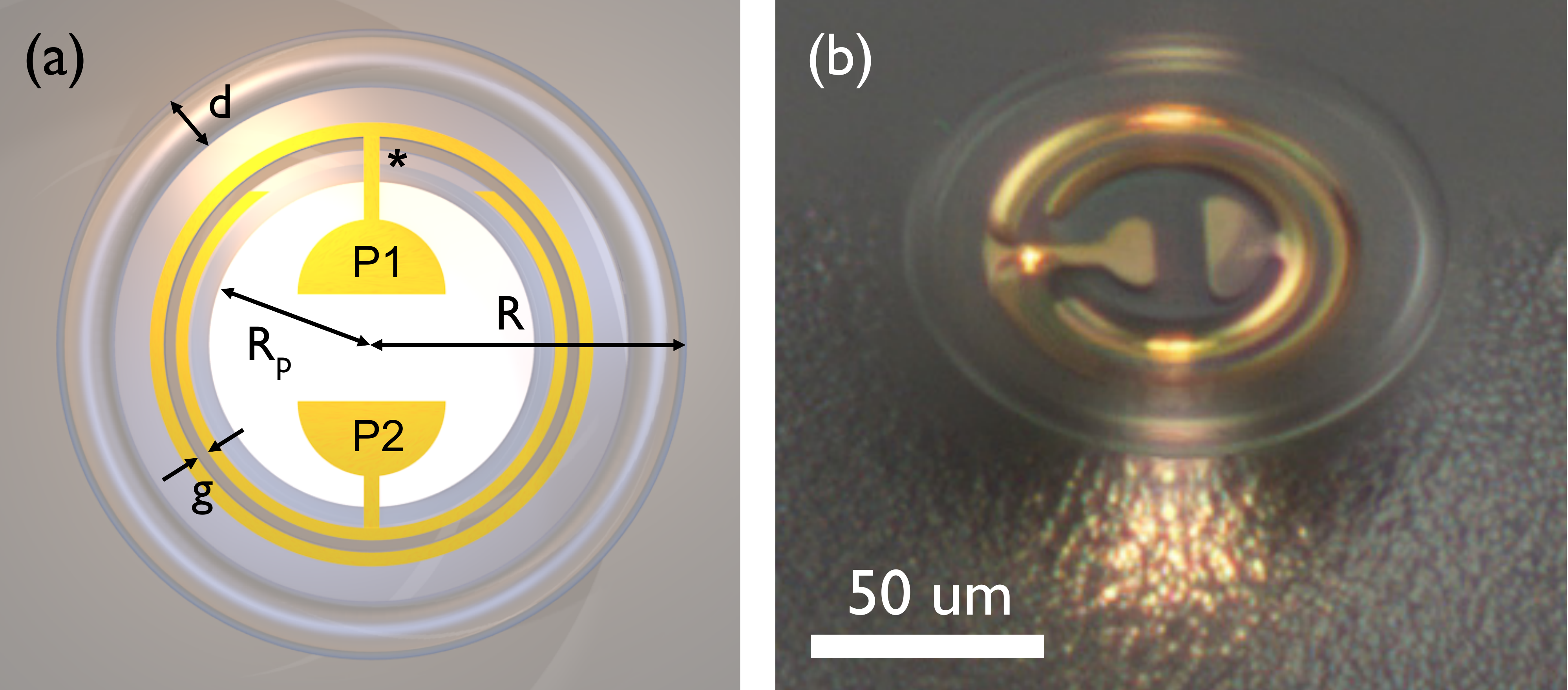}
\caption{Microtoroid based COEMS. (a) Schematic top view of the microtoroid COEMS displaying the relevant dimensions: microtoroid major radius $R$, microtoroid minor diameter $d$, electrode gap $g$ and undercut silicon pedestal radius $R_p$. The gold electrodes have a nominal width of 5 microns. (b) Optical microscope side-view of a fabricated device. This image was obtained by combining 10 individual images taken at different focus to overcome the shallow depth of field (focus stacking).}
\label{Figure1fig}
\end{figure}
Figure \ref{Figure1fig}(a) shows a schematic top view of the designed device.  It consists of a reflown silica microtoroid resonator \cite{armani2003ultra}, in which a circular slot of width $g$ has been etched out, in order to allow for greater mechanical compliance. Gold electrodes are patterned on either side of this slot to enable capacitive actuation.  Upon the application of a bias voltage between the electrodes through the contact pads P1 and P2, the attractive capacitive force strains the microtoroid resulting in an effective reduction in the microtoroid cavity length and an associated frequency shift of the microtoroid whispering gallery resonances. The outer portion of the silica microtoroid is mechanically supported by a single anchor which also serves to connect the outer electrode to the contact pad P1. The position of this anchor is marked by an asterisk in Figure \ref{Figure1fig}(a).
Even with a single narrow anchor, the sag of the outer portion of the microtoroid is minimal as confirmed by finite element method (FEM) simulations. 

\subsection{Modeling}
\label{sectionmodeling}
\begin{figure}
\centering
\includegraphics[width=\textwidth]{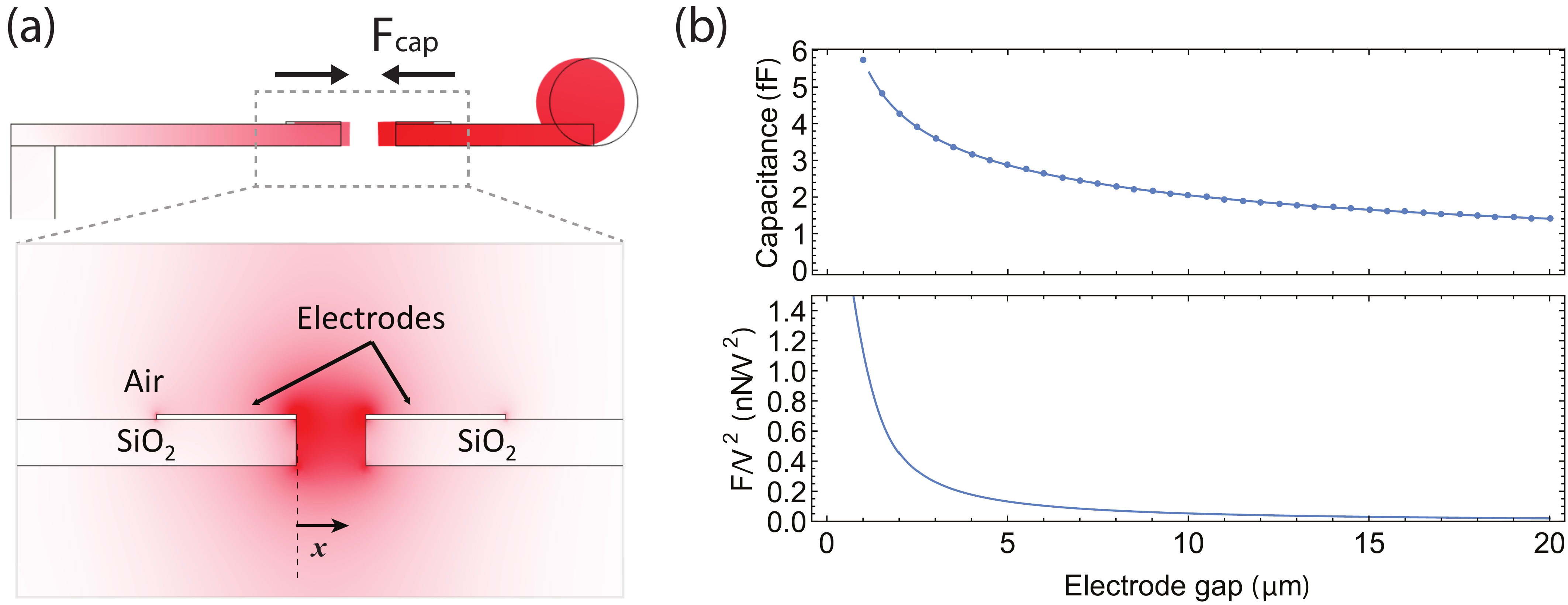}
\caption{(a) Schematic cross section of the microtoroid COEMS. Color code represents the mechanical displacement caused by an applied voltage. Zoomed in view shows the electrodes positioned on either side of the slot etched in the silica. Here the color code represents the electric energy density $\varepsilon E^2$ when a bias voltage is applied. (b) Top panel: calculated capacitance as a function of electrode gap $g$. The dots correspond to the results of individual FEM simulations, while the solid line is a fit to the simulation. Lower panel: Calculated force $F_{\mathrm{cap}}/V^2$, obtained from $C(x)$ through Eq. (\ref{Eqcapacitiveforce}).}
\label{Figuremodeling}
\end{figure}

The application of a bias voltage $V$ between the electrodes leads to an attractive capacitive force $F_{\mathrm{cap}}$: 
\begin{equation}
F_{\mathrm{cap}}=\frac{1}{2}\frac{\mathrm{d}C\left(x\right)}{\mathrm{d}x}\times V^2
\label{Eqcapacitiveforce}
\end{equation}
Here  $x$ is a displacement and $C\left(x\right)$ is the position dependent capacitance formed by the two capacitor plates, see Fig. \ref{Figuremodeling}(a). The value of $C\left(x\right)$ for our device is extracted from FEM simulations and plotted as a function of the gap between the electrodes in the top panel of Fig. \ref{Figuremodeling}(b). The associated force $F_{\mathrm{cap}}/V^2$ given by Eq. (\ref{Eqcapacitiveforce}) is plotted in the lower panel. The  capacitive force leads to a deformation of the outer ring  $\Delta x$ and a shift in the WGM resonance frequencies $\Delta\omega_0$ given by:
\begin{equation}
\Delta\omega_0=g_{\mathrm{om}}\Delta x=g_{\mathrm{om}} \frac{F_{\mathrm{cap}}}{k}\simeq\frac{1}{2 k}\frac{\omega_0}{R}\frac{\mathrm{d}C\left(x\right)}{\mathrm{d}x} V^2\equiv\alpha V^2
\label{Eqtunability}
\end{equation}
where $g_{\mathrm{om}}\simeq \frac{\omega_0}{R}$ is the optomechanical coupling term \cite{baker_photoelastic_2014, bowen2015quantum}, $R$ the major radius of the microtoroid, and $k=m_{\mathrm{eff}}\,\Omega_M^2$ the effective stiffness of the slotted microtoroid which links the effective change in radius of the cavity to the applied force. $\alpha$ is the optical tunability of the structure \cite{pitanti_strong_2015}.
With our nominal fabrication parameters ($R\simeq 60$ microns after reflow, electrode width of 5 microns, gap $g$ of 2 to 3 microns), our modeling predicts a tunability $\alpha/2\pi$ in the range of 3 to 4 kHz/V$^2$.
\subsection{Fabrication}

\begin{figure}
\centering
\includegraphics[width=\textwidth]{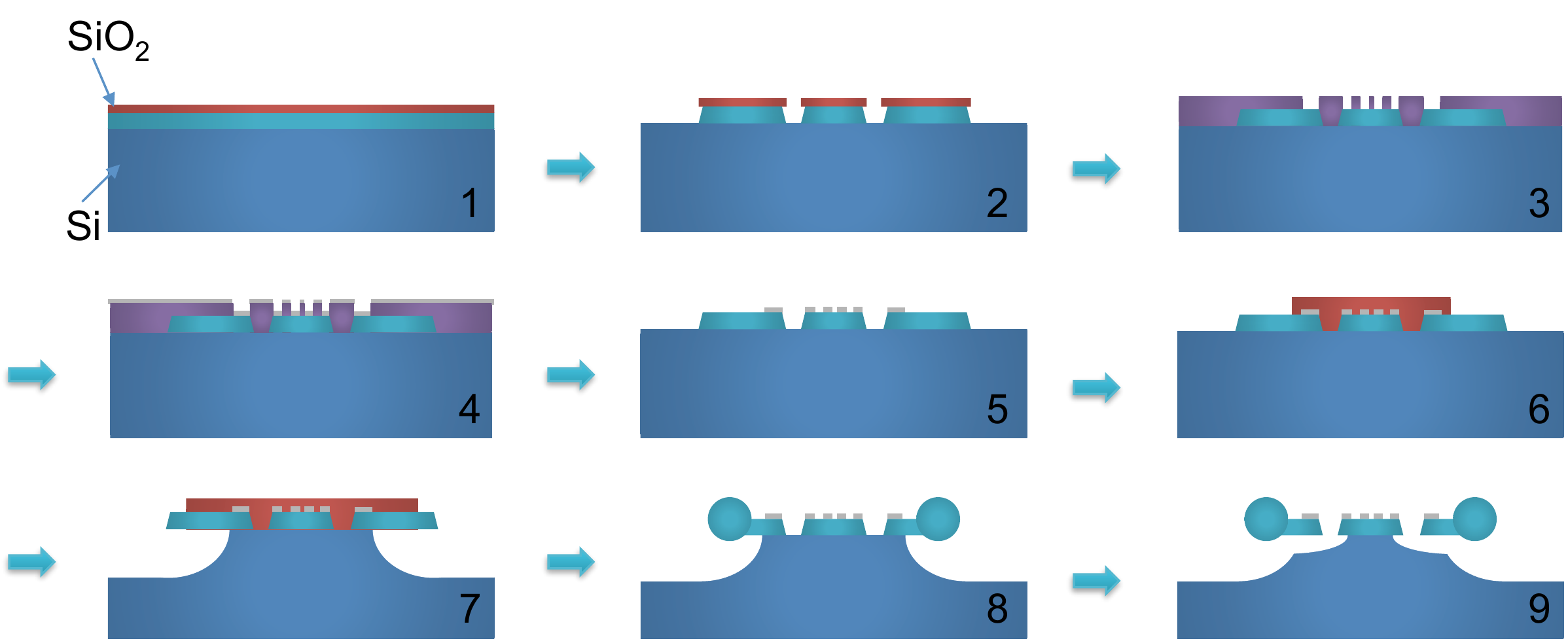}
\caption{Main microfabrication steps for the silica microtoroid COEMS.}
\label{fabricationfig}
\end{figure}

Figure \ref{fabricationfig} outlines the main microfabrication steps. Starting from a silicon wafer with a 2 micron thick SiO$_2$ thermal oxide, slotted disks are patterned with standard photolithography using positive resist (AZ 1518 - red layer), followed by a wet buffered oxide etchant (BOE) etch (steps 1 and 2). The position of the electrodes is defined in step 3, using a mask aligner and negative lift off resist (AZ nLOF 2020 - purple). Next a 10 nm thick tungsten adhesion layer and 100 nm of gold are deposited upon the sample with an e-beam evaporator (step 4) and a lift off is performed by soaking in warm remover (step 5), yielding slotted silica disks with gold electrodes. Steps 6 through 9 describe a two step undercut and silica reflow process, in order to prevent damage to the gold electrodes during reflow. The CO$_2$ laser reflow of silica is a runaway thermal process due to silica's temperature dependent absorption coefficient, whereby increased temperature in the silica results in greater optical absorption and further increases in temperature \cite{armani2003ultra}. However, owing to the high thermal conductivity of silicon,  the silica which is not undercut and remains upon this silicon heat sink remains essentially unaffected by the reflow process.
As a result, the gold electrodes remain unaffected by the CO$_2$ laser pulse provided they are lying over  non-undercut silica. For this reason the central portion of the disk is protected in a third lithography step (step 6), to avoid undercutting through the slot in the first xenon difluoride (XeF$_2$) undercut (step 7). After removal of the protective mask, each disk is exposed to a CO$_2$ laser pulse which reflows the outer rim of the disk, while leaving the electrodes and the narrow anchor pristine (step 8). Finally the chip is further undercut with a second XeF$_2$ underetch step which releases the slot and enables capacitive actuation (step 9). Note that these fabrication steps are similar to those used to fabricate low-loss spoked microtoroid resonators \cite{anetsberger_ultralow-dissipation_2008}. 
Care must be taken to maintain a sufficient distance between the electrodes and the edge of the reflown toroid, in order for the confined optical modes to have sufficiently decayed before they reach the metal and not incur significant loss. FEM simulations (not shown here) show that $\sim$10 microns is sufficient for the WGM electromagnetic energy to have decreased by 70 dB, and can be considered a safe distance.

\section{Characterization}
\label{sectioncharacterization}
To investigate the optical characteristics of the fabricated devices, laser light from a tuneable diode laser in the telecom C band is evanescently coupled into the microtoroid with a pulled fiber taper. Spectroscopy of the devices reveals numerous classes of WGM resonances, separated by a free spectral range of approximately 5 nm. The highest measured quality factors are in the mid $10^7$ range, as shown in Fig. \ref{Figure3fig}(a). Control microtoroids on the same chip fabricated without electrodes were found to have similar Qs, indicative that our current Qs are limited by fabrication and reflow and not by the presence of gold electrodes.
\subsection{DC tunability}
\begin{figure}
\centering
\includegraphics[width=\textwidth]{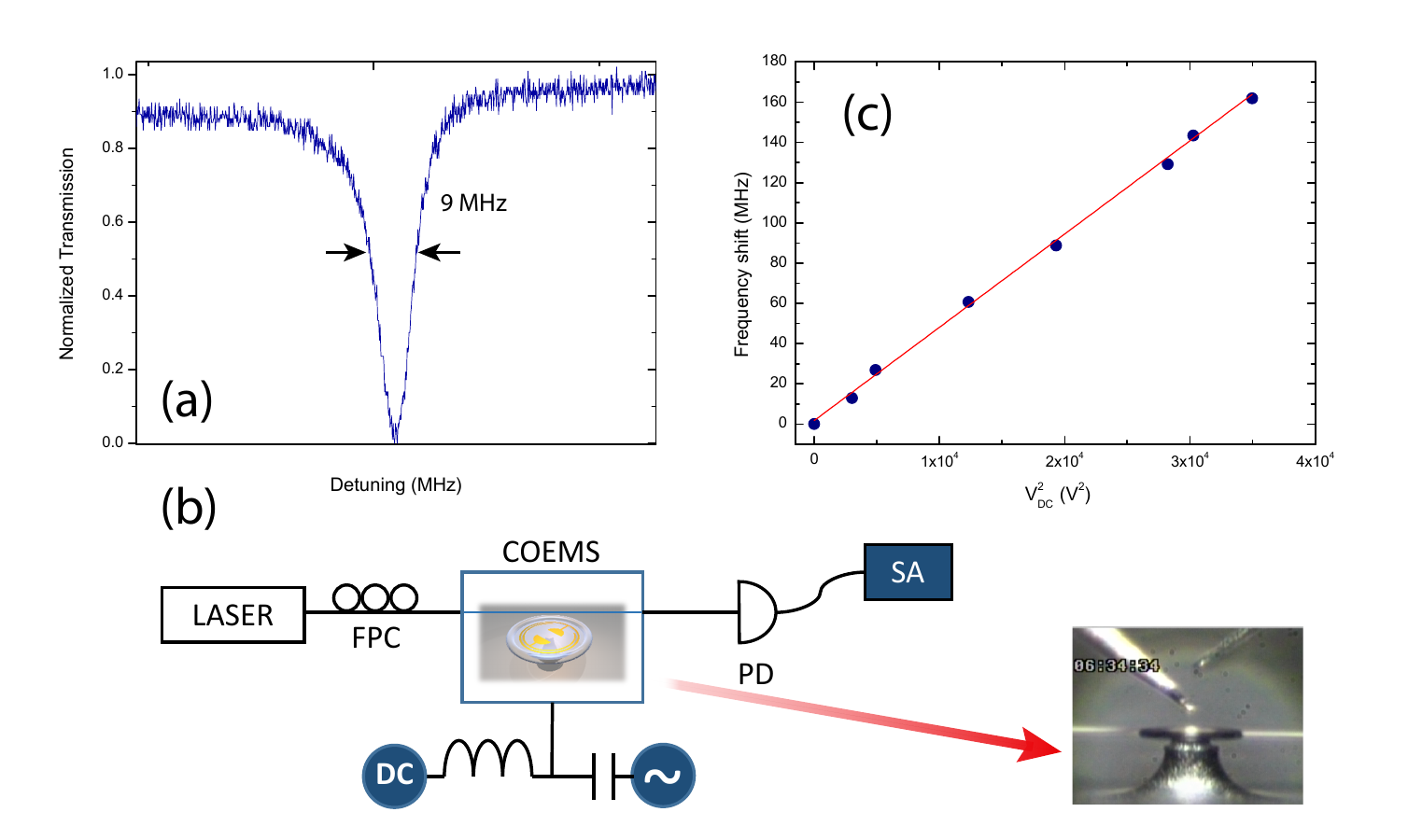}
\caption{(a) Microtoroid WGM resonance near 1550 nm, showing an intrinsic Q factor of $4.5\times 10^7$. (b) Schematic of the experimental setup used to probe the  cavity opto-electromechanical devices. A bias tee allows for the application of both a DC and AC bias on the electrodes. FPC: Fiber polarization controller; PD: photodetector; SA: spectrum analyzer. The inset shows an optical microscope sideview of a microtoroid COEMS positioned in front of the coupling fiber taper, and below two tungsten probe tips. All measurements are performed in air at room temperature. (c) Tunability curve showing optical resonance frequency shift as a function of the square of the applied DC voltage. The red line is a fit to the data. The maximum tuning range reaches approximately 20 linewidths for the largest DC bias.}
\label{Figure3fig}
\end{figure}
Figure \ref{Figure3fig}(b) shows a schematic of the experimental setup used to characterize the response of the devices to an applied voltage across the capacitor plates. The dimensions of the device are too small for the use of conventional wire bonding techniques in the current geometry. Instead, we employ micrometer-sized tungsten probe tips on micropositioning stages to electrically contact the pads P1 and P2 shown in Fig. \ref{Figure1fig}(a). Connection of the electrodes to a bias tee then enables the simultaneous application of DC and AC voltages.  Figure \ref{Figure3fig} (c)  plots the tuning of the device in response to an applied DC bias. Since all WGMs of the microtoroid have a similar optomechanical coupling, Fig. \ref{Figure3fig} (c) can be obtained by tracking the center frequency of any WGM of the microtoroid as the DC bias is gradually ramped up from 0 to 200 V (i.e. as $V_{\mathrm{DC}}^2$ increases from 0 to $4\times 10^4$ V$^2$) and the WGM frequency is gradually increased. The frequency shift exhibits a quadratic dependency with $V_{\mathrm{DC}}$, as expected from Eq. (\ref{Eqtunability}). Fitting the data of Fig. \ref{Figure3fig}(c) yields a tunability $\alpha/2\pi=4.5$ kHz/V$^2$ in reasonable agreement with the results of the FEM simulations discussed in section \ref{sectionmodeling}. We ascribe the somewhat higher value observed in the experiments to the slope in the sidewalls of the slot resulting from the wet etching process, which results in an increased electrode surface area. We show in Fig. \ref{Figure3fig} that the device can be tuned by more than 20 linewidths, enabling switching, routing as well as add/drop applications \cite{klein_reconfigurable_2005}.
Note that the optical Q remains unaltered throughout the tuning procedure, retaining its high value. 
The maximum value of the DC current which can be applied to the electrodes, and therefore the tuning range, is ultimately limited by the breakdown voltage of the electrodes, which was experimentally determined to exceed 300 V and is further discussed in section \ref{sectionbreakdownvoltage}.

\subsection{Broadband operation}

\subsubsection{Harmonic response}
\begin{figure}
\centering
\includegraphics[width=\textwidth]{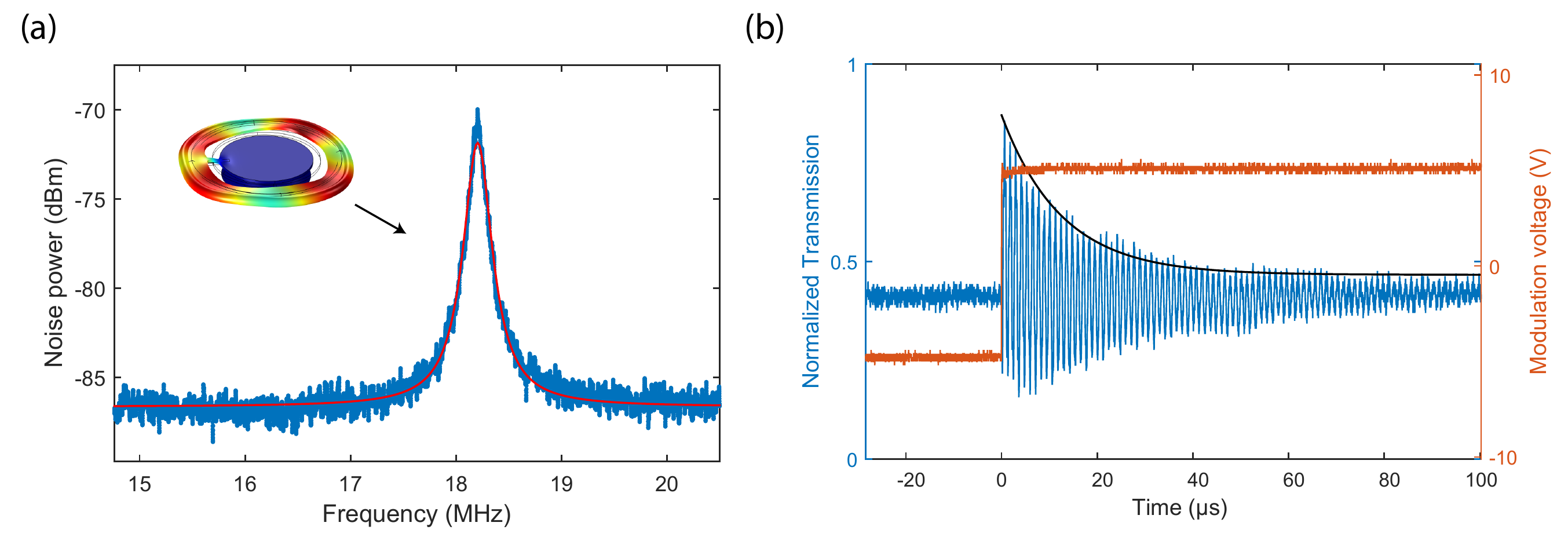}
\caption{ (a) Mechanical "radial breathing mode"-like resonance of the slotted microtoroid near 18 MHz. Blue line: data, red line: lorentzian fit to the data, with fitted mechanical Q $\simeq 180$. FEM simulation of the mechanical displacement profile is shown as an inset.  (b) Transient response of the device to a short duration voltage spike, measured on a WGM with Q $\simeq10^7$. The black line is an exponential fit to the ringdown time. The noise in the normalized transmission is due to noise in the voltage generator and in the diode laser used for this experiment.}
\label{Figbroadbandoperation}
\end{figure}

In this section we focus on the dynamical response of the device, which we will successively characterize according to two main metrics. The first metric corresponds to the response of the device to harmonic driving, and measures how fast the optical output of the device can be harmonically modulated before incurring a cutoff in its response. In the general case, the maximal modulation rate will be limited by one of the following: the capacitor's response time $\tau_C$, the optical cavity lifetime $\tau_{\mathrm{cav}}$ or the mechanical response of the resonator. In our case, because of the small value of C (see Fig. \ref{Figuremodeling}(b)), the capacitor's response time can safely be considered instantaneous. The mechanical response of the resonator is determined by its spectrum of vibrational eigenmodes. A mechanical spectrum of a fabricated device is obtained by detuning the probe laser to the blue side of a WGM resonance of the microtoroid, thus imprinting the mechanical Brownian motion of the resonator onto intensity fluctuations of the photodetected light which are observed with a spectrum analyzer (see Fig. \ref{Figure3fig}(b)). The mechanical spectrum reveals a number of different resonant modes ranging approximately from 1 to 25 MHz.  The mechanical mode corresponding to a `radial breathing mode'-like deformation  of the structure which, because of its essentially in plane and rotationally invariant nature, couples well to both the optical modes of the microtoroid as well as to the capacitive actuation is shown in Fig. \ref{Figbroadbandoperation}(a). Applying concomitantly a DC voltage and  an AC voltage to the electrodes at this mechanical frequency enables efficient modulation of the laser light with a modulation period of 55 ns, three orders of magnitude faster than previously demonstrated thermal modulation schemes on microtoroid resonators \cite{armani2004electrical, heylman_photothermal_2013}. Ultimately, because of the high optical Q of the microtoroids, an upper bound on the modulation frequency is introduced by the finite cavity lifetime $\tau_{\mathrm{cav}}$ of the WGMs. This is in contrast to most previously demonstrated mechanical modulation of lower optical Q cavity schemes \cite{rosenberg_static_2009,  abdulla_tuning_2011,winger_chip-scale_2011,errando2015low}, where the mechanical cutoff at higher frequencies is the limiting factor.
Note that the use of a bias tee, as shown in Fig. \ref{Figure3fig}(b), enables the concomitant application of a large DC bias and small AC modulation. Since the capacitive force scales as $V^2$, this can lead to a greatly increased AC response, providing a 2 V$_{\mathrm{DC}}$/V$_{\mathrm{AC}}$ boost in the frequency shift compared to the no DC case. With a 200 V DC bias, a 5 V peak to peak (Vpp) AC modulation is sufficient to provide full depth (on/off) optical modulation, using the parameters in Fig. \ref{Figure3fig} ($2 V_{\mathrm{DC}} V_{\mathrm{AC}} \alpha/2\pi=9$ MHz), which we observed experimentally.
The voltage requirement for optical modulation is further reduced by a factor $Q_M$ if the AC drive is resonant with a mechanical mode of quality factor $Q_M$. Indeed for an AC drive resonant with the 18 MHz mechanical mode (with Q$_M$=180), only 30 mVpp is required for full depth modulation. 
This very low voltage optical modulation make the capacitively driven microtoroids developed here efficient radio frequency to optical conversion devices \cite{bagci_optical_2014}.

\subsubsection{Transient response}

Next, we consider the transient response of the devices to a voltage spike applied to the electrodes. This spectrally broad pulse will excite many mechanical modes of the microtoroid to various amplitudes,  which then each ring down at their own decay rate. The ringdown time of the device is then a combination of the transient responses of all the excited mechanical modes and is essentially dominated by the ringdown time of the excited mechanical mode with the slowest decay rate. While the previously discussed response to harmonic driving is relevant for some RF to optical conversion applications, this transient response is a pertinent metric for switching and add/drop applications, as it will ultimately limit the rate at which these operations can be realized. 
Figure \ref{Figbroadbandoperation}(b) shows a measurement of the transient response of the microtoroid. A square modulation signal with 10 Vpp and frequency 1 kHz is sent onto the AC port of the bias tee (red trace). The AC port of the bias tee behaves as $\sim$140 kHz high pass filter and converts the square modulation into a series of upwards and downwards voltage spikes with duration $\sim$7 $\mu$s. These spikes excite the radial breathing mode and a combination of lower frequency mechanical modes of the microtoroid, with a resultant effective ringdown time of $\sim 12 \mu$s, corresponding to a frequency of approximately 80 kHz. This value is close to the ringdown time of the RBM (100 kHz). In contrast to harmonic driving where high Q mechanical modes are desirable, for switching applications a low mechanical Q is advantageous to reduce this ringdown time. This could be achieved through engineering of the shape and position of anchors or proper choice of the undercut ratio of the silicon pedestal for example \cite{anetsberger_ultralow-dissipation_2008, nguyen2013ultrahigh}.
Even according to this more stringent and currently unoptimized metric, the  achievable tuning frequencies are nearly two orders of magnitude larger than previously demonstrated techniques \cite{armani2004electrical, heylman_photothermal_2013}.

\section{Improved interdigitated electrode design}

\begin{figure}
\centering
\includegraphics[width=\textwidth]{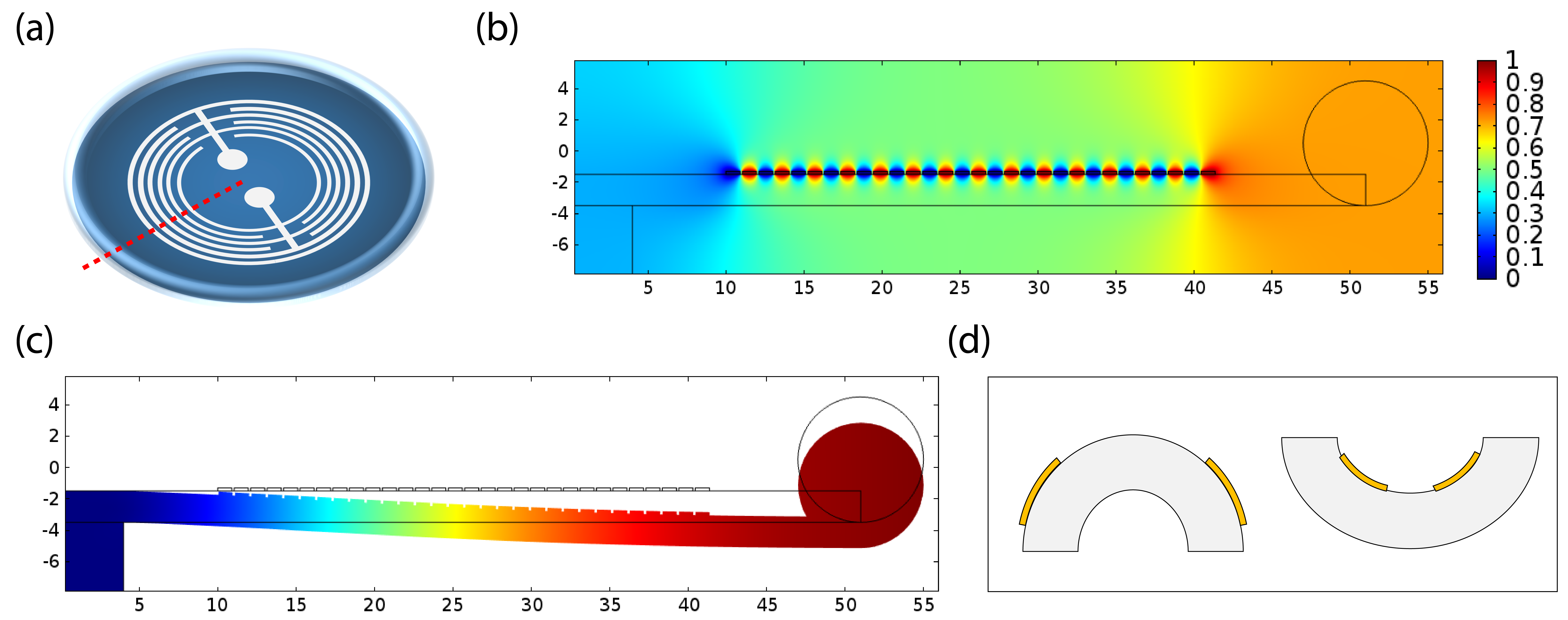}
\caption{Improved microtoroid COEMS design. (a) Schematic illustration of a microtoroid resonator with patterned interdigitated concentric ring electrodes. (b) Cross-sectional view through the dashed red line in (a) revealing N=30 electrodes patterned over the silica disk. Color code represent the electrical potential with electrodes alternatively biased to 0 and 1 V. (c) FEM simulation representing the exaggerated mechanical deformation resulting from an applied bias between the electrodes. The structure deflects downwards and slightly outwards. (d) Schematic illustration of the influence of the dielectric medium under the electrodes. Despite the larger electrode separation, the left case can correspond to a lower energy state because of the higher effective dielectric constant between the electrodes.}
\label{Figurefuture}
\end{figure}

While the fabricated devices already demonstrate efficient optical modulation and fast switching and tuning capabilities, here we theoretically investigate how they could be further improved. In particular increasing the capacitive force $F_{\mathrm{cap}}$ would increase the achievable tuning range and reduce the need for a large DC bias in the experiments. Let us consider for illustrative purposes the case of a parallel plate capacitor, for which analytical solutions for the capacitance and capacitive force exist:
\begin{equation}
C=\frac{\varepsilon_0 K A}{g} \quad \quad
F_{\mathrm{cap}}=-\frac{\varepsilon_0 K A}{2 g^2} V^2
\label{Eqparallelplatecap}
\end{equation}
where $\varepsilon_0=8.85\times10^{-12}$ F m$^{-1}$, $K$, $A$ and $g$ are respectively the vacuum permittivity, dielectric constant of the material between the capacitor plates, the capacitor area and distance between the plates. Equation (\ref{Eqparallelplatecap}) shows that an increase in force per V$^2$ can be achieved through a combination of increased plate area and decreased separation $g$. In the current design the smallest achievable gap size $g$ is limited to a few microns by the resolution of the photolithography and the wet etch step to form the slot in the silica, as shown in Fig. \ref{fabricationfig}. A solution to overcome this limitation is to do away with the slot altogether, as shown in Fig. \ref{Figurefuture}(a).
At the cost of increased mechanical stiffness, removing the slot provides two main advantages. First, it allows for the fabrication of significantly reduced gap sizes through e-beam lithography. Second, it allows for a large increase in the number of electrodes through an interdigitated design, effectively increasing the electrode area in Eq. (\ref{Eqparallelplatecap}). Figure \ref{Figurefuture}(b) shows a cross-sectional view of the proposed approach, with 30 electrodes of width 0.9 micron separated by 150 nm patterned on the top surface. The capacitance calculated for this structure is 250 fF, more than 50 times larger than in our current design. In order to take into account both the effect of the material between the electrodes as well as the fact that the applied force is now distributed throughout the microtoroid surface, we perform full electromechanical FEM simulations to obtain the deformation of the structure in response to an applied bias voltage. The parameters used in this simulation are summarized in Table \ref{Tablerecaptunability}. The results of this simulation, for a 1 V bias applied between the electrodes, are shown in Fig. \ref{Figurefuture}(c). Here the effect of an applied bias is to deform the structure downwards and slightly outwards, effectively increasing the microtoroid's radial dimensions, which is exactly the opposite effect we measured with the slotted microtoroid. This somewhat counter-intuitive behavior can be understood by considering the stored energy inside a capacitor $E_{\mathrm{cap}}=\frac{1}{2}\frac{Q^2}{C}$. For a fixed charge $Q$, the system will seek to minimize its potential energy through an increase in $C$. As shown in Eq. (\ref{Eqparallelplatecap}), this can be achieved by reducing the gap $g$ between the electrodes -as observed previously, but can also be achieved through an increase in the effective dielectric constant $K$ of the material between the electrodes. Here the microtoroid bends downwards (and not upwards as would be expected from a tensile stress being applied to its top surface), so as to maximize the amount of dielectric enclosed between the electrodes. This is shown schematically in Fig. \ref{Figurefuture} (d): provided the dielectric contribution is large enough, the lowest energy solution can actually correspond to an increase in the total separation between the electrodes (left image). To verify this is the correct interpretation, we numerically verify that the downward deflection indeed corresponds to a lower energy state. Furthermore, when adding a high K dielectric coating over the electrodes, the system recovers an upward deflection. With the experimentally achievable parameters of Table \ref{Tablerecaptunability}, this optimized geometry provides a tunability of $\alpha=-0.77$ MHz/V$^2$, a more than two orders of magnitude improvement over current performance, with a bias on the order of 2 V sufficient to displace a high Q WGM resonance by one linewidth.
\begin{table}
\centering
\begin{tabular}{l c c}
\hline
\hline
\textbf{Parameter} & \textbf{Unit} & \textbf{Value}\\
\hline
\hline
Microtoroid major radius $R$ &  $\qquad$ $\mu$m  $\qquad$ &  $\qquad$ 50 $\qquad$\\
\hline
Microtoroid minor diameter $d$ & $\mu$m & 8\\
\hline
Pedestal radius $R_p$ & $\mu$m & 4\\
\hline
Silica layer thickness & $\mu$m & 2\\
\hline
Number of interdigitated electrodes N & - & 30\\
\hline
Electrode width & $\mu$m & 0.9\\
\hline
Electrode gap $g$ & nm & 150\\
\hline
Electrode thickness & nm & 200\\
\hline
tunability $\alpha$ & MHz/V$^2$  & -0.77 \\
\hline
\end{tabular}
\caption{Parameters used in the FEM simulation of the interdigitated microtoroid COEMS design.}
\label{Tablerecaptunability}
\end{table}
Note that here most of the generated deformation is out-of-plane, and therefore not useful as it does not dispersively couple to the microtoroid WGMs. This large out-of plane deflection could however be advantageously leveraged with different resonator geometries \cite{jiang_high-q_2009}.

\section{Conclusion}

We have developed an approach to reversibly tune the optical resonances of  high Q on-chip silica microtoroid resonators based on capacitive actuation, enabled by electrodes patterned on the top surface of the microresonator. This method provides up to tens of MHz modulation rates, several orders of magnitude faster than previously demonstrated heat based techniques. We additionally presented an improved design based on interdigitated electrodes that further improves on current performance. These results add a new level of functionality to a mature platform with a broad range of applications ranging from optomechanics, on chip lasers and bio-sensing to nonlinear optics. Our approach leaves the reflown silica surface intact, and therefore maintains the high optical Q and the possibility to functionalize the silica surface. It may also find use in the active field of RF to optical conversion \cite{bagci_optical_2014, winger_chip-scale_2011, pitanti_strong_2015, andrews_bidirectional_2014}.

\section*{Acknowledgments}

This work was performed in part at the Queensland node of the Australian National Fabrication Facility, a company established under the National Collaborative Research Infrastructure Strategy to provide nano and micro-fabrication facilities for Australia's researchers. This research was funded through the Australian Research Council Linkage grant LP140100595. W.P.B. acknowledges the Australian Research Council Future Fellowship FT140100650.

\section{Appendix}
\subsection{Breakdown voltage}
\label{sectionbreakdownvoltage}

\begin{figure}[h]
\centering
\includegraphics[width=\textwidth]{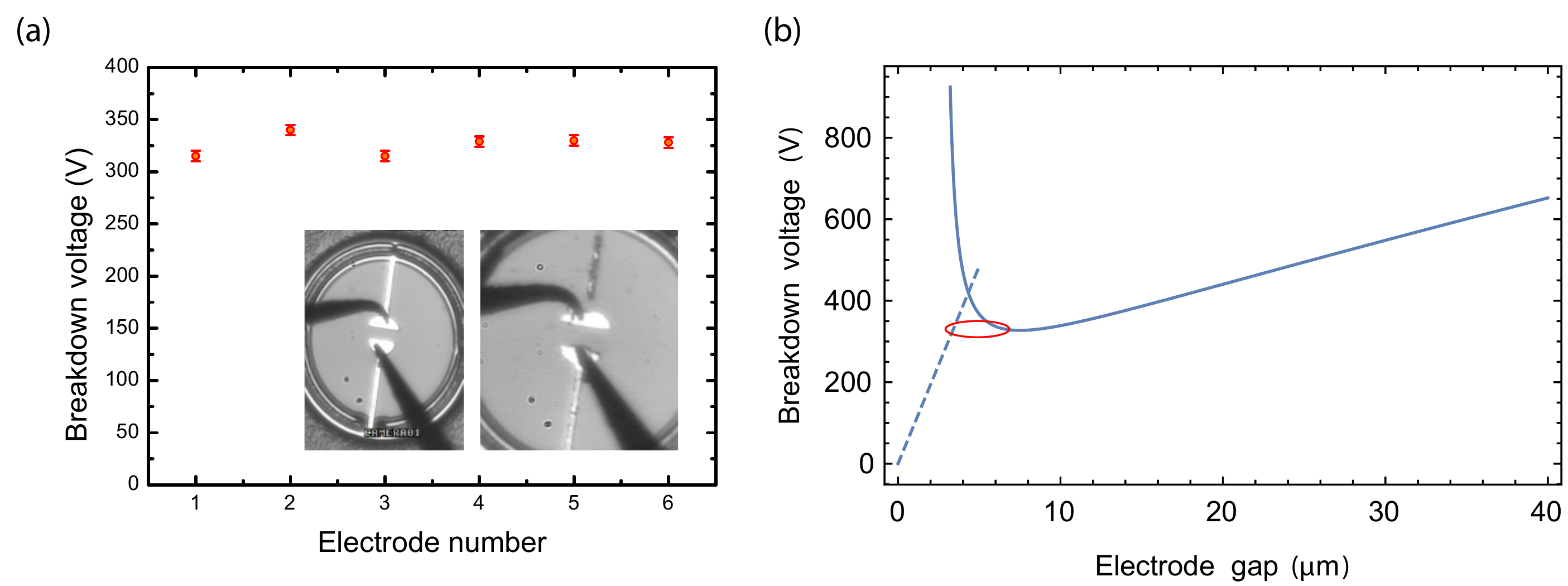}
\caption{(a) Breakdown voltage of the electrodes measured on 6 different devices. At breakdown, the joule heating resulting from the high electrical current density ablates the gold electrodes, which is readily visible through optical microscopy, as shown in the inset. Left image: before breakdown; right image after breakdown. (b) Theoretical breakdown voltage curve in air at 1 atmosphere, as a function of electrode gap distance. The solid line represents the classical Paschen curve, while the dashed blue line is the correction to small gaps for metal electrodes where the breakdown is dominated by electrode vaporization \cite{slade_electrical_2002}. The red ellipse represents the approximate spread of our experimental data.}
\label{breakdownfig}
\end{figure}

The maximum achievable frequency tuning range $\Delta \omega_{\mathrm{max}}$ is limited by the breakdown voltage of the electrodes $V_B$ through $\Delta \omega_{\mathrm{max}}=\alpha V_B^2$ (from Eq. (\ref{Eqtunability})). We experimentally measure the value of $V_B$ on six different devices with gap $g$ ranging from 3 to 7 microns. These results are shown in Fig. \ref{breakdownfig}(a). $V_B$ reproducibly falls inside a band from 310 to 340 volts. Fig. \ref{breakdownfig}(b) shows the theoretical breakdown voltage in air at one atmosphere, as a function of electrode gap distance. It is comprised of two curves; the solid blue curve corresponding to the regime of avalanche ionization of the gas, while the dashed blue line of slope $\sim 100 V/\mu$m corresponds to the regime of field emission current induced vaporization of the metal electrode \cite{slade_electrical_2002} responsible for the trend towards reduced $V_B$ for micrometer and sub-micrometer sized gaps between metal electrodes. Our experimental data is in good agreement with the values expected from the theory.

\end{document}